\documentclass[12pt]{article}
\usepackage{graphicx}

\begin{document}

\title{\textbf{Pion Production in Heavy-ion Collisions in the 1 A GeV region \footnote{Supported by
the National Natural Science Foundation of China under Grant No.
10805061, the special foundation of the president fellowship, the
west doctoral project of Chinese Academy of Sciences, and major
state basic research development program 2007CB815000.}}}

\author{FENG Zhao-Qing$^{1,2} \footnote {Tel: 0931-4969215, 13893620698; Email: fengzhq@impcas.ac.cn}$,
JIN Gen-Ming$^{1,2}$}
\date{}
\maketitle

\begin{center}
$^{1}${\small \emph{Institute of Modern Physics, Chinese Academy of
Sciences, Lanzhou 730000, China}}\\[0pt]
$^{2}${\small \emph{Center of Theoretical Nuclear Physics,
National Laboratory of Heavy Ion Accelerator of Lanzhou,\\[0pt]
Lanzhou 730000, China}}
\end{center}

\begin{abstract}
Within the framework of the improved isospin dependent quantum
molecular dynamics (ImIQMD) model, the pion emission in heavy-ion
collisions in the region 1 A GeV is investigated systematically, in
which the pion is considered to be mainly produced by the decay of
resonances $\triangle$(1232) and N*(1440). The in-medium dependence
and Coulomb effects of the pion production are included in the
calculation. Total pion multiplicity and $\pi^{-}/\pi^{+}$ yields
are calculated for the reaction $^{197}$Au+$^{197}$Au in central
collisions for selected Skyrme parameters SkP, SLy6, Ska, SIII and
compared them with the measured data by the FOPI collaboration.
\end{abstract}

\emph{PACS}: 25.75.-q, 13.75.Gx, 25.80.Ls

\bigskip

Heavy-ion collisions at intermediate energies play a significant
role to extract the information of the nuclear equation of state
(EoS) under extreme conditions, i.e., at high densities and high
temperature. Besides nucleonic observables such as rapidity
distribution and flow, also mesons emitted from the reaction zone
can be probes of the hot and dense nuclear matter, that are also the
interest physics at the Cooling Storage Ring (CSR) energies in
Lanzhou.$^{\cite{Zh08}}$ The emission of pion in heavy-ion
collisions in the region 1 A GeV is especially sensitive as probes
of isospin asymmetric EoS at supra-saturation
densities.$^{\cite{Li08}}$ Spectra of the pion emission in heavy-ion
collisions have been measured by the Kaos and FOPI collaborations
and analyzed systematically by the present theoretical transport
models.$^{\cite{Re07,Mu95}}$ A comparison of the various transport
approaches was made in Ref. $\cite{Ko05}$. The present theoretical
models overpredict the total pion multiplicity if using free
nucleon-nucleon (NN) cross sections below 2 A GeV region compared
with experimental data. Further investigations of the pion emissions
in the 1 A GeV region are still necessary by improving transport
models or developing some new approaches. The improved
isospin-dependent quantum molecular dynamics model has been
successfully applied to treat fusion dynamics and reaction mechanism
of two colliding nuclei near Coulomb
barrier.$^{\cite{Fe05,Fe08,Wa04}}$ To investigate the pion emission,
we further include the inelastic channels in nucleon-nucleon
collisions in the ImIQMD model.

In the ImIQMD model, the time evolutions of the baryons and pions in
the system under the self-consistently generated mean-field are
governed by Hamilton's equations of motion, which read as
\begin{eqnarray}
\dot{\mathbf{p}}_{i}=-\frac{\partial H}{\partial\mathbf{r}_{i}},
\quad \dot{\mathbf{r}}_{i}=\frac{\partial
H}{\partial\mathbf{p}_{i}}.
\end{eqnarray}
Here we omit the shell correction part in the Hamiltonian $H$ as
described in Ref. $\cite{Fe08}$. The Hamiltonian of baryons consists
of the relativistic energy, the effective interaction potential and
the momentum dependent part as follows:
\begin{equation}
H_{B}=\sum_{i}\sqrt{\textbf{p}_{i}^{2}+m_{i}^{2}}+U_{int}+U_{mom}.
\end{equation}
Here the $\textbf{p}_{i}$ and $m_{i}$ represent the momentum and the
mass of the baryons.

The effective interaction potential is composed of the Coulomb
interaction and the local interaction
\begin{equation}
U_{int}=U_{Coul}+U_{loc}.
\end{equation}
The Coulomb interaction potential is written as
\begin{equation}
U_{Coul}=\frac{1}{2}\sum_{i,j,j\neq
i}\frac{e_{i}e_{j}}{r_{ij}}erf(r_{ij}/\sqrt{4L})
\end{equation}
where the $e_{j}$ is the charged number including protons and
charged resonances. The $r_{ij}=|\mathbf{r}_{i}-\mathbf{r}_{j}|$ is
the relative distance of two charged particles. The local
interaction potential is derived directly from the Skyrme
energy-density functional and expressed as
\begin{equation}
U_{loc}=\int V_{loc}(\rho(\mathbf{r}))d\mathbf{r}.
\end{equation}
The local potential energy-density functional reads $\cite{Fe08}$
\begin{equation}
V_{loc}(\rho)=\frac{\alpha}{2}\frac{\rho^{2}}{\rho_{0}}+
\frac{\beta}{1+\gamma}\frac{\rho^{1+\gamma}}{\rho_{0}^{\gamma}}+
\frac{g_{sur}}{2\rho_{0}}(\nabla\rho)^{2}+\frac{g_{sur}^{iso}}{2\rho_{0}}
[\nabla(\rho_{n}-\rho_{p})]^{2}+\frac{C_{sym}}{2\rho_{0}}\rho^{2}\delta^{2}+
g_{\tau}\rho^{8/3}/\rho_{0}^{5/3},
\end{equation}
where the $\rho$ is the baryon density and the
$\delta=(\rho_{n}-\rho_{p})/(\rho_{n}+\rho_{p})$ is the isospin
asymmetry with the proton density $\rho_{p}$ and the neutron density
$\rho_{n}$. The momentum dependent part in the Hamiltonian is
expressed as
\begin{equation}
U_{mom}=\frac{\delta}{2}\sum_{i,j,j\neq
i}\frac{\rho_{ij}}{\rho_{0}}[\ln(\epsilon(\textbf{p}_{i}-\textbf{p}_{j})^{2}+1)]^{2},
\end{equation}
with
\begin{equation}
\rho_{ij}=\frac{1}{(4\pi L)^{3/2}}\exp\left[
-\frac{(\textbf{r}_{i}-\textbf{r}_{j})^{2}}{4L}\right].
\end{equation}
Here the $L$ denotes the square of the pocket-wave width, which is
dependent on the size of the nucleus.

In Table 1 we list the ImIQMD parameters related to several typical
Skyrme forces after including the momentum dependent interaction.
The parameters $\alpha$, $\beta$, $\gamma$, $g_{\tau}$, $g_{sur}$,
$g_{sur}^{iso}$, $\delta$ and $\epsilon$ are related to the Skyrme
parameters $t_{0}, t_{1}, t_{2}, t_{3}$ and $x_{0}, x_{1}, x_{2},
x_{3}$, and determined in order to reproduce the binding energy
($E_{B}$=-16 MeV) of symmetric nuclear matter at saturation density
for a given incompressibility as well as the correct momentum
dependence of the real part of the proton-nucleus optical potential.
In the following calculation we take the Skyrme parameter SLy6,
which can give the good properties from finite nucleus to neutron
star $\cite{Ch97}$.


Analogously to baryons, the Hamiltonian of pions is represented as
\begin{equation}
H_{\pi}=\sum_{i=1}^{N_{\pi}}\left(\sqrt{\textbf{p}_{i}^{2}+m_{\pi}^{2}}+V_{i}^{Coul}\right),
\end{equation}
where the $\textbf{p}_{i}$ and $m_{\pi}$ represent the momentum and
the mass of the pions. The Coulomb interaction is given by
\begin{equation}
V_{i}^{Coul}=\sum_{j=1}^{N_{B}}\frac{e_{i}e_{j}}{r_{ij}},
\end{equation}
where the $N_{\pi}$ and $N_{B}$ is the total number of pions and
baryons including charged resonances. Thus, the pion propagation in
the whole stage is guided essentially by the Coulomb effect. The
in-medium pion potential in the mean field is not considered in the
model. However, the inclusion of the pion optical potential based on
the perturbation expansion of the $\Delta$-hole model gives
negligible influence on the transverse momentum
distribution.$^{\cite{Fu97}}$

The pion is created by the decay of the resonances $\triangle$(1232)
and N*(1440) which are produced in inelastic NN scattering. The
direct pion production cross section is very small in the considered
energies and not included in the model.$^{\cite{Ba01}}$ The reaction
channels are given as follows:
\begin{eqnarray}
NN \leftrightarrow N\triangle, & NN \leftrightarrow NN^{\ast}, & NN
\leftrightarrow \triangle\triangle,
\nonumber \\
\Delta \leftrightarrow N\pi, & N^{\ast} \leftrightarrow N\pi.
\end{eqnarray}
The cross section of each channel to produce resonances are taken
the values calculated with the one-boson exchange
model.$^{\cite{Hu94}}$ Transport models overpredicted the total pion
production with the free cross section. In the ImIQMD model, we use
the free elastic cross section and the in-medium inelastic cross
section which is given by
$\sigma^{inelastic}_{medium}=(\frac{\mu_{BB}^{\ast}}{\mu_{BB}})^{2}\sigma^{inelastic}_{free}$
with the free baryon-baryon (BB) inelastic cross section
$\sigma^{inelastic}_{free}$ and the reduced effective mass
$\mu_{BB}^{\ast}$ (free mass $\mu_{BB}$). The experimental data of
total elastic and inelastic cross sections$^{\cite{Ca93}}$ are
parameterized in the ImIQMD model as shown in Fig.1.


In the 1 A GeV region, there are mostly $\Delta$ resonances which
disintegrate into a $\pi$ and a nucleon, however, the $N^{\ast}$ yet
gives considerable contribution to the high energetic pion yield.
The energy and momentum dependent decay width is used in the ImIQMD
model and expressed as
\begin{equation}
\Gamma(|\textbf{p}|)=\frac{a_{1}|\textbf{p}|^{3}}{(1+a_{2}|\textbf{p}|^{2})(a_{3}+|\textbf{p}|^{2})}\Gamma_{0},
\end{equation}
which originates from the p-wave resonances. The $\textbf{p}$ is the
momentum of the created pion (in GeV/c) in the resonance rest frame.
The values $a_{1}=$22.48 (17.22), $a_{2}=$39.69 and $a_{3}=$0.04
(0.09) are used for the $\Delta$ ($N^{\ast}$) with bare decay width
$\Gamma_{0}=$0.12 GeV (0.2 GeV).$^{\cite{Hu94}}$ In Fig.2 we show a
comparison of the time evolution of the $\pi$, $\Delta$ and
$N^{\ast}$ production in the reaction $^{197}$Au+$^{197}$Au for head
on collisions at 1 A GeV for two cases of the bare decay and energy
dependent decay widths. Both methods almost give the same yield of
the pion production. In the following calculation, we use the energy
and momentum dependent decay width. In Fig.3 we give the
multiplicity of produced pion as a function of the impact parameter
for the same system at 1 A GeV energy. The numbers of produced
$\pi^{-}$, $\pi^{0}$and $\pi^{+}$ are reduced with increasing the
impact parameter because of the decrease of the participants of the
'fire ball' formed in the heavy-ion collisions.


The emission of the produced pion is sensitive to the incident
energy owing to the size of the compressed nuclear matter. We
calculated the transverse momentum distribution of $\pi^{-}$,
$\pi^{0}$and $\pi^{+}$ in central $^{197}$Au+$^{197}$Au collisions
at different incident energies as shown in Fig.4. The larger and
wider distributions were found at the higher incident energies due
to the larger participant numbers of the collision nucleons. The
high energy pions originate from the early phase and the decay of
the N*(1440) resonance also plays a significant
role.$^{\cite{Ma98}}$ In Fig.5 we compare the total pion number and
the $\pi^{-}$/$\pi^{+}$ ratio with the FOPI data in central
$^{197}$Au+$^{197}$Au collisions$^{\cite{Re07}}$ for the Skyrme
parameters SkP, SLy6, Ska and SIII which correspond to the different
incompressibility modulus as listed in table 1. The calculated value
of the total pion number is related to the incompressibility modulus
$K_{\infty}$ and the effective mass in nuclear medium. Over the
whole domain, the force SLy6 is nice and can reproduce the
experimental data. But the parameter slight overpredicts the total
pion multiplicity at lower incident energies and underestimates the
value at higher incident energies if using the above in-medium
inelastic cross section. The in-medium elastic and inelastic cross
sections are still open problems in transport model calculations,
which should be calculated by microscopic many-body models and then
parameterized to add into transport models. The $\pi^{-}$/$\pi^{+}$
ratio is interest for extracting the high density behavior of the
symmetry energy per nucleon.$^{\cite{Li08}}$ Using the isobar model,
one gets the ratio $\pi^{-}$/$\pi^{+}$=1.95 for pions from the
$\Delta$ resonance, and $\pi^{-}$/$\pi^{+}$=1.7 from the $N^{\ast}$
for the system $^{197}$Au+$^{197}$Au.$^{\cite{St86}}$ These
relations are globally valid, i.e. independent of the pion energy.
The observed energy dependence of the $\pi^{-}$/$\pi^{+}$ ratio is
due to the influence of the Coulomb force and the symmetry energy
interaction. The $\pi^{-}/\pi^{+}$ ratio is sensitive to the
stiffness of the symmetry energy at the lower incident energies.
Recently, a soft nuclear symmetry energy at supra-saturation
densities was pointed out by fitting the FOPI data with the IBUU04
model.$^{\cite{Xi09}}$ In the ImIQMD model, we only consider the
linear dependence of the symmetry energy term on the baryon density
as shown in Eq.(6). The inclusion of the density-dependent symmetry
energy in the ImIQMD model is in progress.


In summary, the pion production in heavy-ion collisions in the
region 1 A GeV for the reaction $^{197}$Au+$^{197}$ is investigated
systematically by using the ImIQMD model. The distribution of the
transverse momentum is calculated at different incident energies.
The total number of produced pion and the $\pi^{-}/\pi^{+}$ ratio
are calculated in central collisions for selected Skyrme parameters
SkP, SLy6, Ska, SIII and compared them with the FOPI data.
Deviations from the simple isobar model originate from the Coulomb
and the symmetry interactions. The $\pi^{-}/\pi^{+}$ ratio is
sensitive to the stiffness of the symmetry energy at the lower
incident energies that may be further investigated at the CSR
energies.

We would like to thank Prof. Lie-Wen Chen, Prof. Wei Zuo and Dr.
Gao-Chan Yong for fruitful discussions.

\newpage
\begin{table}
\caption{ImIQMD parameters and properties of symmetric nuclear
matter for Skyrme effective interactions after the inclusion of the
momentum dependent interaction with parameters $\delta$=1.57 MeV and
$\epsilon$=500 c$^{2}$/GeV$^{2}$} \vspace*{-10pt}
\begin{center}
\def\temptablewidth{0.8\textwidth}
{\rule{\temptablewidth}{1pt}}
\begin{tabular*}{\temptablewidth}{@{\extracolsep{\fill}}ccccccccc}
&Parameters         &SkM*  &Ska  &SIII  &SVI  &SkP  &RATP  &SLy6 \\
\hline
&$\alpha$ (MeV)     &-325.1 &-179.3 &-128.1 &-123.0 &-357.7 &-250.3 &-296.7 \\
&$\beta$  (MeV)     &238.3  &71.9   &42.2   &51.6   &286.3  &149.6  &199.3 \\
&$\gamma$           &1.14   &1.35   &2.14   &2.14   &1.15   &1.19   &1.14 \\
&$g_{sur}$(MeV fm$^{2}$) &21.8 &26.5  &18.3 &14.1   &19.5   &25.6   &22.9 \\
&$g_{sur}^{iso}$(MeV $fm^{2}$) &-5.5 &-7.9 &-4.9 &-3.0 &-11.3 &0.0 &-2.7 \\
&$g_{\tau}$ (MeV)        &5.9  &13.9  &6.4  &1.1    &0.0   &11.0    &9.9 \\
&$C_{sym}$ (MeV)         &30.1 &33.0  &28.2 &27.0   &30.9  &29.3    &32.0 \\
&$\rho_{\infty}$ (fm$^{-3}$) &0.16 &0.155 &0.145 &0.144 &0.162 &0.16 &0.16 \\
&$m_{\infty}^{\ast}/m$ &0.639 &0.51  &0.62 &0.73 &0.77 &0.56 &0.57 \\
&$K_{\infty}$ (MeV)    &215   &262   &353  &366  &200  &239  &230 \\
\end{tabular*}
{\rule{\temptablewidth}{1pt}}
\end{center}
\end{table}

\begin{figure}
\begin{center}
{\includegraphics*[width=0.8\textwidth]{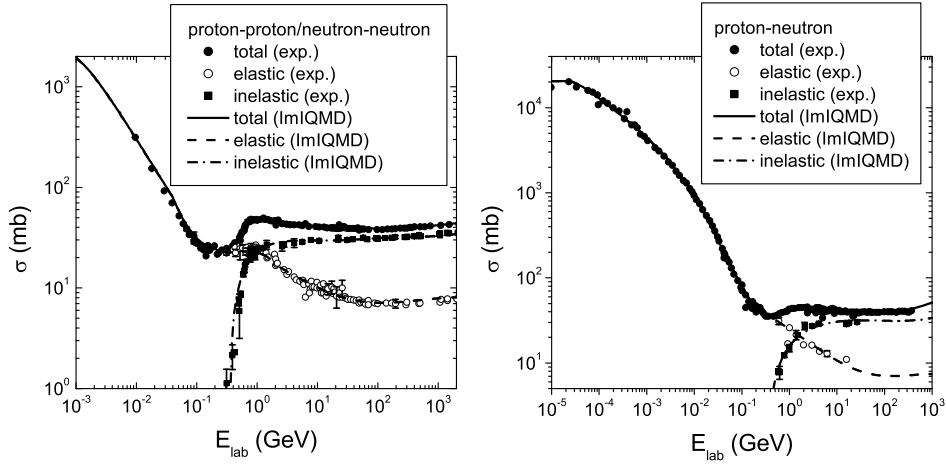}}
\end{center}
\caption{Comparison of nucleon-nucleon cross sections parameterized
in ImIQMD and the experimental data.$^{\cite{Ca93}}$}
\end{figure}

\begin{figure}
\begin{center}
{\includegraphics*[width=0.8\textwidth]{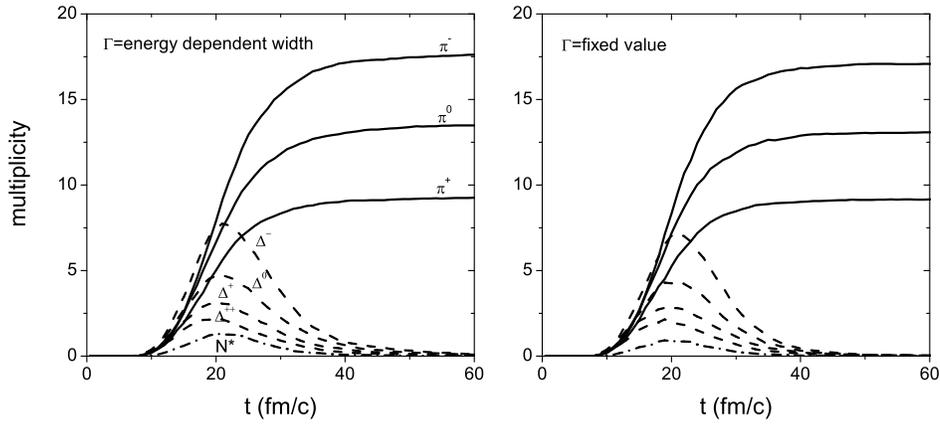}}
\end{center}
\caption{Production of pion, delta and N* for head-on collisions in
the reaction $^{197}$Au+$^{197}$Au at 1 A GeV as functions of
evolution time with the energy dependent decay width (left panel)
and fixed width (right panel).}
\end{figure}

\begin{figure}
\begin{center}
{\includegraphics*[width=0.8\textwidth]{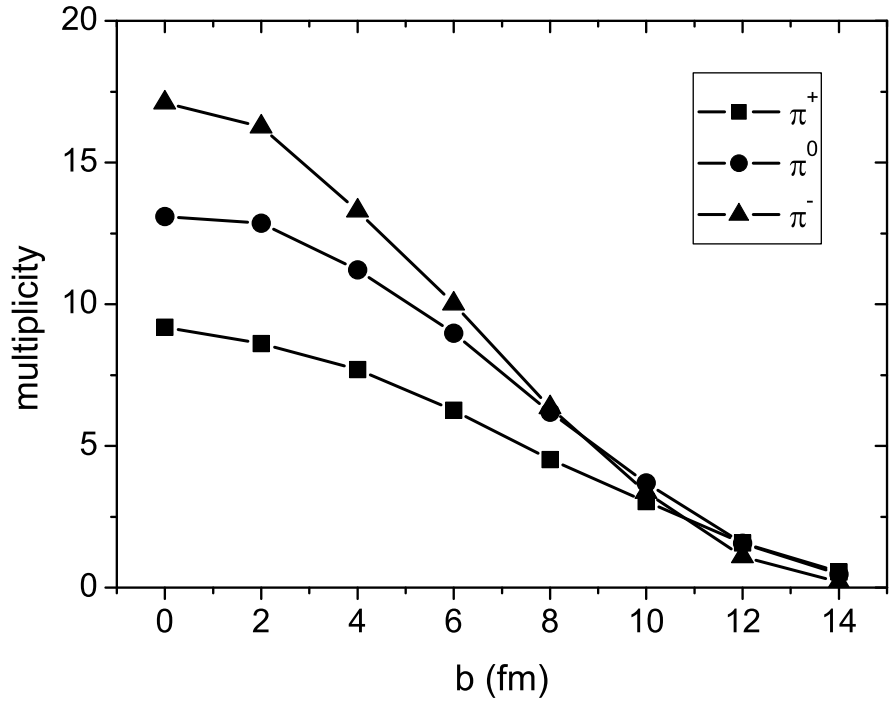}}
\end{center}
\caption{Final multiplicities of $\pi^{-}$, $\pi^{0}$ and $\pi^{+}$
as a function of impact parameter for head-on collisions in the
reaction $^{197}$Au+$^{197}$Au at 1 A GeV.}
\end{figure}

\begin{figure}
\begin{center}
{\includegraphics*[width=0.8\textwidth]{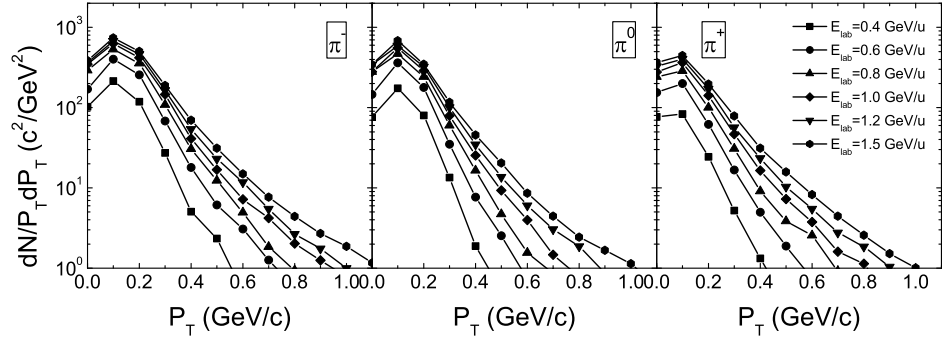}}
\end{center}
\caption{Final transverse momentum distribution in central
$^{197}$Au+$^{197}$Au collisions at different incident energies.}
\end{figure}

\begin{figure}
\begin{center}
{\includegraphics*[width=0.8\textwidth]{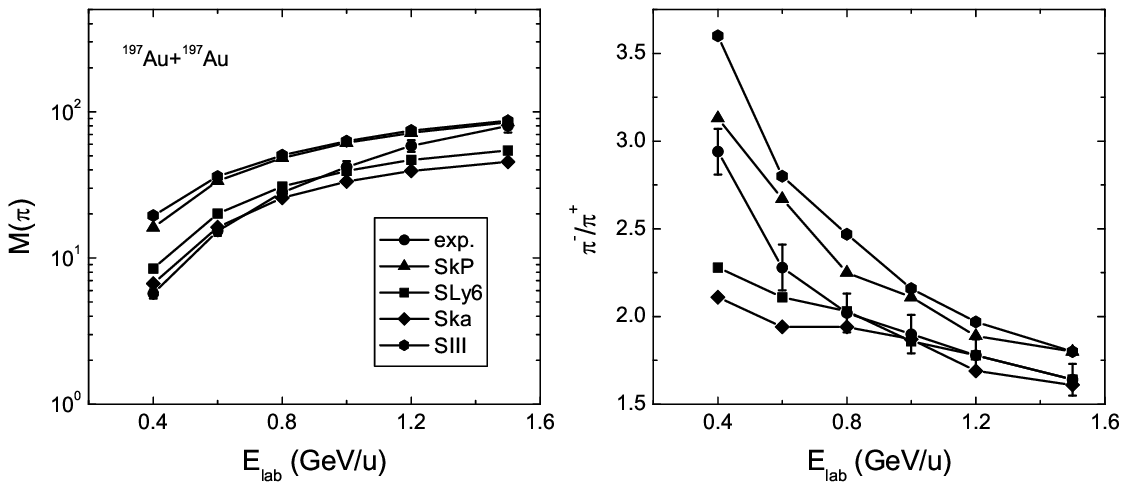}}
\end{center}
\caption{Calculated excitation functions of the total pion
multiplicity (left panel) and the ratio $\pi^{-}$/$\pi^{+}$ (right
panel) in central $^{197}$Au+$^{197}$Au collisions and compared with
the FOPI data.$^{\cite{Re07}}$}
\end{figure}


\begin{thebibliography}{99}
\bibitem{Zh08} Zhan W L, Xia J W, Zhao H W et al (HIRFL-CSR Group) 2008 \emph{Nucl. Phys. A}
\textbf{805} 533c
\bibitem{Li08} Li B A, Chen L W, Ko C M 2008 \emph{Phys. Rep.} \textbf{464}
113
\bibitem{Re07} Reisdorf W, Stockmeier M, Andronic A et al (FOPI collaboration) 2007 \emph{Nucl. Phys. A} \textbf{781}
459
\bibitem{Mu95} C. M\"{u}ntz et al (KaoS collaboration) 1995 \emph{Z. Phys. A} \textbf{352}
175
\bibitem{Ko05} Kolomeitsev E E, Hartnack C, Barz H W et al 2005 \emph{J. Phys. G} \textbf{31}
S741
\bibitem{Fe05} Feng Z Q, Zhang F S, Jin G M, Huang X 2005 \emph{Nucl. Phys. A}
\textbf{750} 232
\bibitem{Fe08} Feng Z Q, Jin G M, Zhang F S 2008 \emph{Nucl. Phys.
A} \textbf{802} 91
\bibitem{Wa04} Wang N, Li Z X, Wu X Z et al 2004 \emph{Phys. Rev. C} \textbf{69} 034608.
\bibitem{Ch97} Chabanat E, Bonche P, Haensel P et al 1997 \emph{Nucl.
Phys. A} \textbf{627} 710
\bibitem{Fu97} Fuchs C, Sehn L, Lehmann E et al 1997 \emph{Phys. Rev. C} \textbf{55} 411.
\bibitem{Ba01} Li B A, Sustich A T, Zhang B, Ko C M 2001 \emph{Int. J Mod. Phys.
E} \textbf{10} 1
\bibitem{Ca93} Catherine L-L, Fran\c{c}ois L 1993 \emph{Rev. Mod.
Phys.} \textbf{65} 47
\bibitem{Hu94} Huber S, Aichelin J 1994 \emph{Nucl. Phys. A} \textbf{573} 587
\bibitem{Ma98} Maheswari V S, Fuchs C, Faessler A et al 1998 \emph{Nucl. Phys. A} \textbf{628}
669
\bibitem{St86} Stock R 1986 \emph{Phys. Rep.} \textbf{135} 259
\bibitem{Xi09} Xiao Z G, Li B A, Chen L W et al 2009 \emph{Phys. Rev.
Lett.} \textbf{102} 062502

\end{thebibliography}
\end{document}